\documentstyle[12pt]{article}
\catcode`\@=11
\long\def\@makefntext#1{
\protect\noindent \hbox to 3.2pt {\hskip-.9pt
$^{{\ninerm\@thefnmark}}$\hfil}#1\hfill}		%CAN BE USED

\def\@makefnmark{\hbox to 0pt{$^{\@thefnmark}$\hss}}  %ORIGINAL

\def\ps@myheadings{\let\@mkboth\@gobbletwo
\def\@oddhead{\hbox{}
\rightmark\hfil\ninerm\thepage}
\def\@oddfoot{}\def\@evenhead{\ninerm\thepage\hfil
\leftmark\hbox{}}\def\@evenfoot{}
\def\sectionmark##1{}\def\subsectionmark##1{}}

\renewcommand{\thefootnote}{\fnsymbol{footnote}}

\newcounter{sectionc}\newcounter{subsectionc}\newcounter{subsubsectionc}
\renewcommand{\section}[1] {\vspace*{0.6cm}\addtocounter{sectionc}{1}
\setcounter{subsectionc}{0}\setcounter{subsubsectionc}{0}\noindent
	{\normalsize\bf\thesectionc. #1}\par\vspace*{0.4cm}}
\renewcommand{\subsection}[1] {\vspace*{0.6cm}\addtocounter{subsectionc}{1}
	\setcounter{subsubsectionc}{0}\noindent
	{\normalsize\it\thesectionc.\thesubsectionc. #1}\par\vspace*{0.4cm}}
\renewcommand{\subsubsection}[1]
{\vspace*{0.6cm}\addtocounter{subsubsectionc}{1}
	\noindent {\normalsize\rm\thesectionc.\thesubsectionc.\thesubsubsectionc.
	#1}\par\vspace*{0.4cm}}

\newcounter{appendixc}
\newcounter{subappendixc}[appendixc]
\newcounter{subsubappendixc}[subappendixc]

\renewcommand{\appendix}[1] {\vspace*{0.6cm}
        \refstepcounter{appendixc}
        \setcounter{figure}{0}
        \setcounter{table}{0}
        \setcounter{equation}{0}
        \renewcommand{\thefigure}{\Alph{appendixc}.\arabic{figure}}
        \renewcommand{\thetable}{\Alph{appendixc}.\arabic{table}}
        \renewcommand{\theappendixc}{\Alph{appendixc}}
        \renewcommand{\theequation}{\Alph{appendixc}.\arabic{equation}}
        \noindent{\bf Appendix \theappendixc #1}\par\vspace*{0.4cm}}

\def\abstracts#1{{

\centering{\begin{minipage}{12.2truecm}\footnotesize\baselineskip=12pt\noindent
	\centerline{\footnotesize ABSTRACT}\vspace*{0.3cm}
	\parindent=0pt #1
	\end{minipage}}\par}}

\renewenvironment{thebibliography}[1]
	{\begin{list}{\arabic{enumi}.}
	{\usecounter{enumi}\setlength{\parsep}{0pt}
\setlength{\leftmargin 1.25cm}{\rightmargin 0pt}
	 \setlength{\itemsep}{0pt} \settowidth
	{\labelwidth}{#1.}\sloppy}}{\end{list}}

\newcounter{itemlistc}
\newcounter{romanlistc}
\newcounter{alphlistc}
\newcounter{arabiclistc}

\newcommand{\fcaption}[1]{
        \refstepcounter{figure}
        \setbox\@tempboxa = \hbox{\footnotesize Fig.~\thefigure. #1}
        \ifdim \wd\@tempboxa > 6in
           {\begin{center}
        \parbox{6in}{\footnotesize\baselineskip=12pt Fig.~\thefigure. #1}
            \end{center}}
        \else
             {\begin{center}
             {\footnotesize Fig.~\thefigure. #1}
              \end{center}}
        \fi}

\newcommand{\tcaption}[1]{
        \refstepcounter{table}
        \setbox\@tempboxa = \hbox{\footnotesize Table~\thetable. #1}
        \ifdim \wd\@tempboxa > 6in
           {\begin{center}
        \parbox{6in}{\footnotesize\baselineskip=12pt Table~\thetable. #1}
            \end{center}}
        \else
             {\begin{center}
             {\footnotesize Table~\thetable. #1}
              \end{center}}
        \fi}

\def\@citex[#1]#2{\if@filesw\immediate\write\@auxout
	{\string\citation{#2}}\fi
\def\@citea{}\@cite{\@for\@citeb:=#2\do
	{\@citea\def\@citea{,}\@ifundefined
	{b@\@citeb}{{\bf ?}\@warning
	{Citation `\@citeb' on page \thepage \space undefined}}
	{\csname b@\@citeb\endcsname}}}{#1}}

\newif\if@cghi
\def\cite{\@cghitrue\@ifnextchar [{\@tempswatrue
	\@citex}{\@tempswafalse\@citex[]}}
\def\citelow{\@cghifalse\@ifnextchar [{\@tempswatrue
	\@citex}{\@tempswafalse\@citex[]}}
\def\@cite#1#2{{$\null^{#1}$\if@tempswa\typeout
	{IJCGA warning: optional citation argument
	ignored: `#2'} \fi}}

 1
 1
 1

\font\ninerm=cmr9

\def\a{\alpha}
\def\b{\beta}

\def\s{\sigma}

\def\o{\over}
\begin{document}

\centerline{\normalsize\bf STRINGS IN COSMOLOGICAL SPACETIMES AND
THEIR BACK-REACTION}
\baselineskip=22pt
\baselineskip=16pt
%\vfill
%\vspace*{0.6cm}
\centerline{\footnotesize H. J. DE VEGA}
\baselineskip=13pt
\centerline{\footnotesize\it Laboratoire de Physique Th\'eorique et
Hautes Energies,}
\centerline{\footnotesize\it Universit\'e Pierre et Marie Curie (Paris VI) et
Universit\'e Denis
Diderot (Paris VII),}
\centerline{\footnotesize\it  Tour 16, 1er. \'etage, 4, Place Jussieu}
\baselineskip=12pt
\centerline{\footnotesize\it
75252 Paris, cedex 05, France.
Laboratoire Associ\'{e} au CNRS URA280.}
%\vfill
\vspace*{0.9cm}
\begin{center}
Lecture delivered at STRINGS '95, Future Perspectives in String Theory,
USC, Los Angeles, march 13 - 18, 1995.
\end{center}
\abstracts{ This is a short review on strings in curved spacetimes. We
start by recalling the classical and quantum string behaviour  in
singular plane waves backgrounds.
We then report on  the string behaviour in cosmological spacetimes
(FRW, de Sitter, power inflation) which is by now largerly understood.
Recent progress on self-consistent solutions to the Einstein equations
for string dominated universes is reviewed. The energy-momentum
tensor for a gas of strings is  considered  as source of the
spacetime geometry. The string
equation of state is determined from the behaviour of the explicit
string solutions.
This yields a  self-consistent cosmological
  solution    exhibiting  realistic matter dominated behaviour
$ R \sim (T)^{2/3}\; $ for large times and  radiation dominated
 behaviour $ R \sim (T)^{1/2}\; $ for early times.
Inflation in the string theory context is discussed.}

%\vspace*{0.6cm}
\normalsize\baselineskip=15pt
\setcounter{footnote}{0}
\renewcommand{\thefootnote}{\alph{footnote}}
\section{Strings and Quantum Gravity}

	 The construction of a sensible quantum theory of gravitation is
 probably the greatest challenge in today's theoretical physics.
	 Deep conceptual problems (as the lost of quantum coherence)
arise when one tries to
 combine (second) quantization  concepts with General Relativity. That
is, it may be very
 well that a quantum theory of gravitation needs new concepts and ideas.
 Of course, this future theory must have the today's General Relativity and
 Quantum Mechanics (and QFT) as limiting cases. In some sense, what everybody
  is doing in this domain (including string theories approach) may be
  to the real
 theory what the old quantum theory in the 10's was compared with
 quantum mechanics \cite{eri92}.

	 The main drawback to develop a quantum theory of gravitation is
 clearly the total lack of experimental guides for the theoretical
 development. The physical effects combining
 gravitation and quantum mechanics are relevant only at energies of the
 order of  $M_{Planck}  =  \hbar c / G  =  1.22 \,  10^{16} $Tev.
 Such energies were available in the Universe at times $ t < t_{Planck}
 =  5.4 \, 10^{-44} $sec. Anyway, as a question of principle,
 to construct a quantum theory of gravitation is a problem of fundamental
  relevance for theoretical physics . In addition, one cannot rule
 out completely the possibility of some ``low energy'' ($E \ll  M_{Planck}$)
 physical effect that could be experimentally tested.

   Since  $M_{Planck}$ is the heaviest possible particle scale, a theory valid
 there (necessarily involving quantum gravitation) will also be valid at any
lower
 energy scale. One may ignore higher energy phenomena in a low energy theory,
  but not the opposite. In other words, it will be a `theory of everything'.
 We think that this is the {\bf key point} on the quantization of gravity.
 A theory that holds till the Planck scale must describe {\bf all}
what happens
 at lower energies including all known particle physics as well as what
  we do not know yet (that is, beyond the standard model).
 Notice that this conclusion is totally independent of the use of
string models.
  A direct important consequence of this conclusion, is that it does not
  make physical sense to quantize {\bf pure gravity}. A physically
 sensible quantum theory cannot contain only gravitons. To give an example,
  a theoretical prediction for graviton-graviton scattering at energies of
 the order of $M_{Planck}$   must include all particles produced in a real
 experiment. That is, in practice, all existing particles in nature, since
 gravity couples to all matter\cite{eri92}.

String theory is  a serious candidate for a quantum
description of gravity since it provides a unified model of all
interactions overcoming at the same time the nonrenormalizable
character of quantum fields theories of gravity.

As a first step on the understanding of quantum gravitational
 phenomena in a string framework, we started in 1987 a programme of string
 quantization on curved spacetimes \cite{dvs87,negro}.

The investigation of
strings in curved spacetimes is currently the best framework to study the
physics of gravitation in the context of string theory since it
provides essential
clues about the physics in this context but is clearly not the end of
the story. The next step  beyond the investigation of {\bf test}
strings, consist in finding {\bf self-consistently} the geometry from
the strings as matter sources for the Einstein equations
or better the string effective equations (beta functions).
This goal is achieved in \cite{cos} for cosmological spacetimes at the
classical level. Namely, we used  the energy-momentum tensor for a gas
of strings as source for the Einstein equations and we solved them
self-consistently. [For more detailed reviews see \cite{eri92,eri94}].

\bigskip

Let us consider bosonic strings (open or closed) propagating in a
curved D-dimensional spacetime defined by a metric $G_{AB}(X),
0 \leq A,B \leq D-1$.
  The action can be written as
 \begin{equation}\label{accion}
    S  = {{1}\o{2 \pi \a'}} \int d\s d\tau \sqrt{g}\,  g_{\a\b}(\s,\tau) \;
 G_{AB}(X) \,
 \partial^{\a}X^A(\s,\tau) \, \partial^{\b}X^B(\s,\tau)
 \end{equation}
 Here  $ g_{\a\b}(\s,\tau)$  ( $0 \leq \a, \b \leq 1$ ) is the metric in the
 worldsheet, $\a'$ stands for the string tension. As in flat spacetime, $\a'
 \sim (M_{Planck})^{-2} \sim ( l_{Planck})^2$  fixes the scale in the theory.

We will start considering given gravitational backgrounds  $G_{AB}(X)$.
That is, we start to investigate {\em test} strings propagating on a
given spacetime. In section 3,
the back reaction problem will be studied. That is, how the strings
may act as source of the geometry.

String propagation in massless backgrounds other
than gravitational (dilaton, antisymmetric tensor) can be investigated
analogously.

The string equations of motion and constraints follow by extremizing
eq.(\ref{accion}) with respect to $X^{A}(\s,\tau)$ and  $
g_{\a\b}(\s,\tau)$, respectively.
In the conformal gauge, they take the form:
 \begin{eqnarray}\label{conouno}
 \partial_{-+}X^{A}(\s,\tau)  +   \Gamma^{A}_{BC}(X)\,  \partial_{+}
  X^{B}(\s,\tau) \, \partial_{-}X^{C}(\s,\tau) =  0~, \quad
   0 \le A \le D-1 ,
\end{eqnarray}
\begin{eqnarray} \label{conodos}
         T_{\pm\pm} \equiv G_{AB}(X) \, \partial_{\pm}X^{A}(\s,\tau) \,
\partial_{\pm}X^{B}(\s,\tau)= 0\, , \quad  T_{+-} \equiv T_{-+} \equiv 0
  \end{eqnarray}
 where we introduce light-cone variables  $x_{\pm} \equiv \s \pm \tau $
   on the world-sheet and where  $\Gamma^{A}_{BC}(X)$  stand for the
 connections (Christoffel symbols) associated to the metric  $ G_{AB}(X)$.

Notice that these equations in the conformal gauge are still invariant
under the conformal reparametrizations:
 \begin{equation}\label{conftr}
 \s +  \tau \to  \s'  +  \tau' =  f(\s+ \tau) \qquad , \qquad
\s - \tau \to \s' - \tau'   =g(\s-\tau)
 \end{equation}
Here $f(x)$ and  $g(x)$ are arbitrary functions.

The string boundary conditions in curved spacetimes are identical to
 those in Minkowski spacetime. That is,
 \begin{eqnarray}\label{condc}
  X^{A}(\s + 2 \pi,\tau) = \,   X^{A}(\s,\tau) \quad & {\rm closed
\,strings} \cr \cr
    \partial_{\s}X^{A}(0,\tau)=\, \partial_{\s} X^{A}(\pi,\tau) =  \,0
 \quad &  {\rm open \,strings. }
 \end{eqnarray}

We shall consider, as usual, that only four space-time dimensions are
uncompactified. That is, we shall consider the strings as living
on the tensor  product of a curved four dimensional space-time with
lorentzian signature and a compact space which is there  to cancel
the anomalies. Therefore, from now on strings will propagate in the
curved (physical) four dimensional space-time. However, we will find
instructive to study the case where this curved space-time has
dimensionality  $D$, where $D$ may be 2, 3 or arbitrary.

 \section{ Strings Falling into  Spacetime Singularities:   nonlinear
plane waves.}

 Let us  first  consider strings propagating in
 gravitational plane-wave space-times \cite{ondpl,ondpl2}.   In this
geometry the full
 non-linear string equations  (\ref{conouno}) and constraints
(\ref{conodos}) can be   exactly solved in closed form.
 The plane-wave space-times are  described by the metric
\begin{equation}\label{opla}
 (ds)^2  =  - dU dV + \sum_{i=1}^{D-2}(dX^{i})^2  - \left[ W_{1}(U)\,
(X^2 - Y^2 )
 + 2 \,W_{2}(U) \,X Y \right]\, (dU)^2
\end{equation}
 where $ X \equiv X^1$ , $Y \equiv X^2$ .  These space-times are exact
 solutions of the vacuum Einstein equations for any choice of the
 profile functions $ W_{1}(U)$ and $W_{2}(U)$.
  In addition they are exact string vacua \cite{guven}.
  The case when $W_{2}(U) = 0$  describes waves of constant
 polarization. When both $ W_{1}(U) \neq 0$ and  $W_{2}(U)\neq 0$ ,
 eq.(\ref{opla}) describes waves with arbitrary  polarization.
  If $W_{1}(U)$ and/or  $W_{2}(U)$ are singular functions, space-time
 singularities will be present. The singularities will be located on
 the null plane  $U =$ constant. We consider profiles which are nonzero
 only on a finite interval  $-T < U < T$ , and which have power-type
  singularities \cite{ondpl,ondpl2},
 \begin{equation}\label{singu}
 W_{1}(U) \buildrel{U \to 0}\over= { {\a_{1}} \o {|U|^{\beta_{1}}}}
\quad , \quad
 W_{2}(U)  \buildrel{U \to 0}\over= { {\a_{2}} \o {|U|^{\beta_{2}}}}
 \end{equation}
 The spacetimes (\ref{opla}) share many properties with the shockwaves
 \cite{camas,horo}. In particular, $U(\s,\tau)$ obeys the d'Alembert equation
 and we can choose the light-cone gauge
\begin{equation}\label{gauge}
 U = 2 \, \a' p^U \tau \; .
\end{equation}
The string equations
 of motion (\ref{conouno}) become then in the metric (\ref{opla}) :
 \begin{eqnarray}\label{equac}
 V''  -  \ddot V ~ + ~(2 \a' p^U )^2 \left[ \partial_{U} W_{1}\,(X^2 - Y^2 )
 ~ +~2 \,\partial_{U}W_{2} \,X Y \right]\cr
  + ~8 \a' p^U \left[ W_{1} ( X \dot X - Y \dot Y ) +W_{2} (X \dot Y +
\dot Y X )
 \right] =  \,0  \cr
 X'' - \ddot X ~ +~ (2 \a' p^U )^2 \left[ W_{1} X -W_{2} Y \right] = \,0
 \cr Y'' - \ddot Y ~ +~ (2 \a' p^U )^2 \left[ W_{2} X -W_{1} Y \right] = \, 0
\end{eqnarray}
 and the constraints (\ref{conodos}) take the form:
 \begin{equation} \label{vinop}
 \pm \partial_{\pm}V_{<}  =   {1 \o {\a' p^{U}}}
 \left[ (\partial_{\pm}X)^2 +
 (\partial_{\pm}Y)^2 + \sum_{i=3}^{D-2}(\partial_{\pm}X^{i})^2
  \right] + \a' p^{U} \left[  W_{1}\, (X^2 - Y^2 )
 + 2 \,W_{2} \,X Y \right]
  \end{equation}
 Let us analyze now the solutions of the string equations
(\ref{equac}) and (\ref{vinop})
 for a closed string. The transverse coordinates obey the d'Alembert
equation, with the solution
 \begin{eqnarray} \label{stra}
X^{i}(\s,\tau)  =  q^i + 2 p^i \a' \tau + i \sqrt{\a'} \sum_{n \neq 0}
   \{ \a^{i}_{n} \exp[in(\s - \tau)]
+ \tilde  \a^{i}_{n} \exp[-in(\s + \tau )] \}/n , \cr ~ 3 \leq i
\leq D-2  . \quad
\end{eqnarray}
 For the $X$ and $Y$ components it is convenient to Fourier expand as
 $$
 X(\s,\tau) =  \sum_{n=-\infty}^{+\infty} \exp(in\s) ~ X_{n}(\tau)\quad , \quad
 Y(\s,\tau) =  \sum_{n=-\infty}^{+\infty} \exp(in\s) ~ Y_{n}(\tau)
 $$
 Then, eqs.(\ref{equac}) for $X$ and $Y$ yield
\begin{eqnarray}\label{schr}
  \ddot X_{n} + n^2 X_{n}-(2 \a' p^U )^2
 \left[ W_{1} X_{n} - W_{2} Y_{n} \right]= ~ 0 \cr
  \ddot Y_{n} + n^2 Y_{n}-(2 \a' p^U )^{2}
 \left[ W_{2} X_{n} - W_{1} Y_{n}\right]= ~ 0
\end{eqnarray}
 where we consistently set $ U  = 2 \a' p^U \tau$.
Formally, these are two coupled one dimensional Schr\"odinger-like
 equations with $\tau$ playing the r\^ole of a spatial coordinate.

 We study now the interaction of the string with the gravitational
 wave. For  $2\a' p^{U} \tau < -T$, $W_{1,2}(\tau) = 0$ and therefore
$X, Y$ are  given by the usual flat-space expansions
 $$
 X(\s,\tau)  =~  q^X_{<} + 2 p^X_{<} \a' \tau + i \sqrt{\a'} \sum_{n \neq 0}
   \{ \a^{X}_{n<} \exp[-in\tau)]
	 - \tilde  \a^{X}_{-n<} \exp[in\tau )] \} \exp[in\s]/n
$$
$$
 Y(\s,\tau)  =~  q^Y_{<} + 2 p^Y_{<} \a' \tau + i \sqrt{\a'} \sum_{n \neq 0}
   \{ \a^{Y}_{n<} \exp[-in\tau)]
	 - \tilde  \a^{Y}_{-n<} \exp[in\tau )] \} \exp[in\s]/n
% \eqn\inso
 $$
 These solutions define the initial conditions for the string
 propagation in $\tau \geq -\tau_{0} \equiv -{T \o {2\a' p^{U}}}$.
 In the language of the Schr\"odinger-like equations we have a
 two channel potential in the interval $-\tau_{0} < \tau < +\tau_{0}$.
 We consider the propagation of the string when it approaches the
 singularity at $U = 0 = \tau$ from $\tau < 0$.

The general case when $W_1 \neq 0 \neq W_2$ is solved in
\cite{ondpl2}. Let us concentrate here on the case $W_2 \equiv 0$,
$W_1(U) = \a \left [ |U|^{-\b} -   |T|^{-\b} \right]$ for $ |U| < T$.

 Eq.(\ref{schr}) can be approximated near $\tau = 0^-$ as
 $$
 \ddot X_{n}-{{(2 \a' p^U )^{2-\beta}}\o{|\tau|^{\beta}}}
  \; \a \; X_{n} = 0
$$
$$
 \ddot Y_{n}+{{(2 \a' p^U )^{2-\beta}}\o{|\tau|^{\beta}}}
 \; \a  \; Y_{n}= 0
 $$
 The behaviour of the solutions $X_{n}(\tau)$ and $Y_{n}(\tau)$
 for $\tau \to 0$ depends crucially on the value range of $\beta$. Namely,
 i) $\beta > 2$,
 ii) $\b = 2$,
 iii) $\beta < 2$.

%For simplicity, we start
 When $\beta < 2$ the solution for $\tau\to 0^{-}$ behaves as
 $$
 X(\s, \tau)\buildrel{\tau\to 0^{-}}\over = B^X(\s) + A^X(\s)~ \tau +
O(|\tau|^{2-\beta}) \; , \;
$$
$$
 Y(\s,\tau)\buildrel{\tau\to 0^{-}}\over =  B^Y(\s) + A^Y(\s)~ \tau +
O(|\tau|^{2-\beta})~~,~~\beta \neq 1
 $$
 [In the special case $\beta = 1$ one should add a term  $0 ( \tau
\ln|\tau|)$]. Here and in what follows, $  B^X(\s), A^X(\s), B^Y(\s)$
and $ A^Y(\s)$ are arbitrary functions depending on the initial data.

For  $\beta < 2$, the string coordinates $X, Y$ are always regular
 indicating that the string propagates smoothly through the
 gravitational-wave singularity U = 0. (Nevertheless, the
 velocities $\dot X$ and $\dot Y$ diverge at $\tau = 0$ when $1 \leq
\beta < 2$).

For the case $\beta = 2$  the solution is \cite{ondpl}:
\begin{eqnarray}\label{beta2}
 X(\s,\tau)\buildrel{\tau\to 0^{-}}\over =  B^X(\s)
|\tau|^{(1-\sqrt{1+4\a})/2} +  A^X(\s)
|\tau|^{(1+\sqrt{1+4\a})/2} \cr
Y(\s,\tau)\buildrel{\tau\to 0^{-}}\over =  B^Y(\s)
|\tau|^{(1-\sqrt{1-4\a})/2} +  A^Y(\s)
|\tau|^{(1+\sqrt{1-4\a})/2}
\end{eqnarray}

 Let us now consider the case $\beta > 2$. We have \cite{ondpl}
 \begin{eqnarray}\label{betaG}
 X(\s,\tau)&\buildrel{\tau\to 0^{-}}\over=& B^X(\s)
 |\tau|^{{\beta}\o 4} ~ \exp[K|\tau|^{1-\beta/2}]  +
A^X(\s)
 |\tau|^{{\beta}\o 4} ~ \exp[-K|\tau|^{1-\beta/2}]    \quad ,
 \quad \cr
Y(\s,\tau)&\buildrel{\tau\to 0^{-}}\over=& A^Y(\s)
 |\tau|^{{\beta}\o 4} ~ \cos{\left[K |\tau|^{1-\beta/2} + C^Y(\s)
\right]} \; .
\end{eqnarray}
where
$$
K = \,{{(2 \a' p^{U} )^{1-\beta/2}}\o
 {\beta/2 - 1}}\sqrt{\a} > 0 \quad , \quad
 $$

Let us now analyze the string behaviour near
 the singularity $\tau \to 0^- $ for $\beta \geq 2$. We see that for
strong enough singularities ($\b \leq 2$) the transverse coordinate
$X$ tends to infinity when the string approaches the singularity $\tau
\to 0, U\to 0$. This means that the string does not cross the
gravitational wave, since it does not reach the $U>0$ region. The $Y$
coordinate tends to zero oscillating, when $\tau \to 0$.

The string goes off to $X=\infty$,
 grazing the singularity plane $U = 0$ (therefore never crossing it).
 At the same time, the string oscillates in the $Y$ direction,
with an amplitude vanishing
 for $\tau \to 0^- $.

	 The spatial string coordinates $X^{i}(\s,\tau)~ [3 \leq i \leq D-2]$
 behave freely [eq.(\ref{stra})].
	 The longitudinal coordinate $V(\s,\tau)$ follows from the
 constraint eqs.(\ref{vinop}) and the solutions (\ref{beta2},\ref{betaG})
for $X(\s,\tau),
 Y(\s,\tau)$ and  $X^{j}(\s,\tau)~ [3 \leq j \leq D-2]$. We see that
 for $\tau \to 0^- $ , $V(\s,\tau)$
 diverges as the square of the singular solutions  (\ref{beta2},\ref{betaG}).

   Let us consider the spatial length element of the string, i.e. the
 length at fixed  $U = 2 \a'p^{U} \tau $  , between  two points
 $(\s,\tau)$  and  $(\s+d\s,\tau)$,
 $$
 ds^2 = dX^2 + dY^2 + \sum_{j=3}^{D-2} (dX^{j})^2
 $$
 For $\tau\to 0^{-}$ eqs.(\ref{beta2},\ref{betaG})  yield
 \begin{eqnarray}\label{explo}
 ds^2\buildrel{\tau\to 0^{-}}\over =
 ~  \left[ {B^{X}}(\s)'\right]^{2} d\s^2 ~~ |\tau|^{1-\sqrt{1+4\a}}
\quad  {\rm for}  ~
 \beta=2 \, , \cr
 ds^2\buildrel{\tau\to 0^{-}}\over =
 ~  \left[ {B^{X}}(\s)'\right]^{2}
d\s^2 ~~ |\tau|^{\beta/2} ~ \exp\left[K|\tau|^{1-\beta/2}\right]
 \quad  {\rm for} ~ \beta \geq 2.
\end{eqnarray}
 That is, the proper length between $(\s_{0},\tau)$ and
$(\s_{1},\tau)$ is given by
 \begin{eqnarray}\label{strech}
 \Delta s \buildrel{\tau\to 0^{-}}\over =&
 [B^X(\s_{1}) - ~B^X(\s_{2})]~ \sqrt{|\tau|}^
 {1-\sqrt{1+4\tilde\a}}\to \infty \quad {\rm for} ~ \beta=2 \, , \cr
 ~\Delta s \buildrel{\tau\to 0^{-}}\over =&
 [B^X (\s_{1}) - ~B^X(\s_{2})]~~ |\tau|^{\beta/4} ~
 \exp\left[K|\tau|^{1-\beta/2}\right] \to \infty
  ~{\rm  for} ~ \beta \geq 2.
\end{eqnarray}
 We see that  $\Delta s \to \infty$ for $\tau\to 0^{-}$ . That is,
 the string stretches infinitely when it approaches the singularity plane.
 This stretching of the string proper size also occurs for $\tau \to 0$
 in the inflationary cosmological backgrounds as we shall see below.

	 Another consequence of eqs. (\ref{beta2},\ref{betaG})  is that the
 string reaches infinity in a finite time $\tau$. In particular, for
 $\s$-independent coefficients, eqs. (\ref{beta2},\ref{betaG})
describe geodesic
 trajectories. The fact that for  $\beta \geq 2$, a point particle
 (as well as a string) goes off to infinity in a finite $\tau$ indicates
 that the space-time is singular.

  Finally, we would like to remark that the string evolution near the
space-time
 singularity is a {\bf collective motion} governed by the nature of the
 gravitational field. The (initial) state of the string fixes the overall
 $\s$-dependent coefficients
 $A^X(\s),B^X(\s),$ $ A^Y(\s), B^Y(\s)$ [see
eqs. (\ref{beta2},\ref{betaG})], whereas the
 $\tau$-dependence is fully determined by the space-time geometry. In other
 words, the $\tau$-dependence is the same for all modes $n$. In some
 directions, the string collective propagation turns to be an
 infinite motion (the escape direction $X$), whereas in the orthogonal
direction ($Y$),
 the motion is oscillatory, but with a fixed ($n$-independent) frequency.
  In fact, these features are not restricted to singular gravitational
  waves, but {\bf are generic} to strings in strong gravitational
 fields [see sec.(6) and refs.(\cite{camas,eri92})].

 For sufficiently weak spacetime singularities
  ($\beta_{1} < 2$ and $\beta_{2} < 2$), the string crosses the
 singularity and reaches the region $U > 0$. Therefore, outgoing
 scattering states and outgoing operators can be defined in the region
 $U > 0$. We explicitly found in \cite{ondpl,ondpl2} the transformation
 relating the ingoing and
 outgoing string mode operators. For the particles described by the
 quantum string states, this relation implies two types of effects
 as described in \cite{trans,eri92} for generic asymptotically flat
spacetimes: (i)
  rotation of spin polarization in the $(X,Y)$ plane, and (ii)
{\bf  transmutation} between different particles. We computed in
\cite{ondpl,ondpl2}
  the expectation values of the outgoing mass  $M_{>}^{2}$  operator
 and of the mode-number operator $N_{>}$, in the ingoing ground state
$|O_{<}>$.
 As for shockwaves (see  \cite{camas} ) , $M_{>}^{2}$  and $N_{>}$ have
 different expectation values  than  $M_{<}^{2}$  and  $N_{<}$ .
  This difference is due to the excitation of the string modes after
 crossing the space-time singularity. In other words, the string state
 is not an eigenstate of $M_{>}^{2}$, but an infinity superposition
 of one-particle states with different masses. This is a consequence
 of the particle transmutation
 which allows particle masses different from the initial one.

\section{Strings and Multistrings in Cosmological Spacetimes and the
Self-consistent string cosmology}

Recently, several interesting progresses in the understanding of string
propagation in cosmological spacetimes have been made
\cite{prd}${}^-$\cite{ijm}. The classical string equations of motion plus
the string constraints were shown to be exactly integrable in
D-dimensional de Sitter spacetime, and equivalent to a
Toda-type  model with a potential unbounded from below. In 2+1
dimensions, the string dynamics in de Sitter spacetime is exactly
described by the sinh-Gordon equation.

{\bf Exact} string solutions were systematically found by soliton
methods using the linear system associated to the problem
(the so-called dressing method in soliton theory)
\cite{dms,cdms}. In addition, exact circular string solutions
were found in terms of elliptic functions\cite{dls}. All these solutions
describe one string, several strings or even an infinite number of
different and independent strings. A single world-sheet simultaneously
describes many different strings. This is a new feature appearing as a
consequence of the interaction of the strings with the spacetime geometry.
Here, interaction among the strings (like splitting and merging) is neglected,
the only interaction is with the curved background. Different types of
behaviour appear in the multistring solutions. For some of them the
energy and proper size are bounded (`stable strings') while for many
others the energy and size blow up for large radius of the universe
($R \to \infty$, `unstable strings'). In
addition, such stable and  unstable string behaviours are exhibited by
the ring solutions found in \cite{din} for
Friedmann-Robertson-Walker (FRW) universes and for power type inflationary
backgrounds. In all these works, strings were considered as {\it test}
objects propagating on the given {\it fixed} backgrounds.

We report here  the  recent results \cite{cos} further in the
investigation of the physical properties of the string solutions above
mentioned. We compute the energy-momentum tensor of these strings and
we use it to
find the back reaction effect on the spacetime. That is, we investigate
whether these classical strings can  sustain the corresponding cosmological
background. This is achieved by considering {\bf self-consistently},
the strings as matter sources for the Einstein (general relativity)
equations (without the dilaton field), as well as for the
string effective equations (beta functions) including the dilaton, the
dilaton potential and the central charge term.

 In spatially homogeneous and isotropic universes,
\begin{equation}
ds^2 = (dT)^2 - R(T)^2 \sum_{i=1}^{D-1}(dX^i)^2
\label{metA}
\end{equation}
the string energy-momentum tensor $T_A^B(X) \; , (A,B=1,\ldots D) $
for our string solutions takes the fluid form, allowing us to define
the string pressure $p$ through $\, -\delta_i^k~p = T_i^k \, $ and the
string energy density as $ \rho = T_0^0 $. The continuity equation
$D^A\;T_A^B = 0 $ takes then  the form
\begin{equation}
\dot{\rho} + (D-1)\, H\, (p + \rho ) = 0 ,
 \label{contA}
\end{equation}
where
$ H  \equiv  {{d\log R}\over{dT}} $.
We consider $D=1+1, D=2+1$, and generic $D-$dimensional universes.

In $1+1$ cosmological spacetimes we find the general solution of the
string equations of motion and constraints for arbitrary expansion
factor $ R $ . It consists of two families: one depends on two
arbitrary functions $ f_{\pm}(\sigma \pm \tau) $ and has {\it constant}
energy density $ \rho $ and  {\it negative} pressure $
p = -  \rho $. That is, a perfect fluid relation holds
\begin{equation}
p = (\gamma - 1 ) \rho
\label{flup}
\end{equation}
with $\gamma = 0$ in $ D = 1 + 1 $ dimensions. The other family of
solutions depends on {\it two} arbitrary constants and describes a
massless point particle (the string center of mass). This second
solution has $  p = \rho =  u \; R^{-2} > 0 $ . This is a perfect
fluid type relation with $\gamma = 2 $.
These behaviours fulfil the continuity equation (\ref{contA}) in $ D = 2$.

In $2+1$ dimensions and for any factor $ R $ , we find that circular
strings exhibit three different asymptotic behaviours :
\begin{itemize}
\item (i) {\bf unstable} behaviour for $ R \to \infty $   in inflationary
universes (this
corresponds to conformal time $ \eta \sim \tau \to \tau_0 $ with
 finite $ \tau_0 $ and proper
string size $ S \sim R \to \infty $), for which the string energy $ E_u
\sim R \to \infty $ and the string pressure $ p_u \simeq -E_u/2 \to -\infty
$ is {\bf negative} . This behavior dominates for  $ R \to \infty $ in
inflationary universes.
\item (ii) {\bf Dual} to unstable behaviour for $ R \to 0 $. This
corresponds to  $ \eta \sim (\tau - \tau_0)^{-1} \to +\infty $ for
finite $ \tau  \to \tau_0 $, $ S \sim R \to 0 $
(except for de Sitter spacetime
where $ S \to 1/H $ ), for which the string energy $ E_d \sim 1/R \to
\infty $ and the string pressure $ p_d \simeq E/2 \to + \infty $ ,
is {\bf positive}.
\item (iii) {\bf Stable} for $ R \to \infty $, (corresponding to $\eta
\to \infty , \tau\to \infty, S = $ constant ), for which the string
energy is $ E_s = $ constant and the string pressure vanishes $ p_s = 0 $ .
\end{itemize}

Here the indices $(u,d,s)$ stand for `unstable', `dual' and `stable'
respectively. The behaviours (i) and (ii) are related by the duality
transformation  $ R \leftrightarrow 1/R $ , the case (ii) being
invariant under duality. In the three cases, we find perfect fluid relations
(\ref{flup}) with the values of $\gamma$ :
\begin{equation}
\gamma_u = 1/2 \quad , \quad \gamma_d = 3/2 \quad , \quad\gamma_s = 1~.
\label{glup}
\end{equation}
For a perfect gas of strings on a comoving volume $ R^2 $, the energy
density $ \rho $ is proportional to $ E / R^2 $, which yields the
scaling  $ \rho_u = u \, R^{-1} , \rho_d = d \, R^{-3} , \rho_s = s\, R^{-2}$.
All densities and pressures obey the continuity equation (\ref{contA})
as it must be.

The $ 1+1 $ and $ 2+1 $ string solutions here described exist in any
spacetime dimension. Embedded in D-dimensional universes, the
 $ 1+1 $ and $ 2+1 $  solutions  describe straight strings and
circular strings, respectively. In D-dimensional spacetime,
 strings may  spread in $ D - 1 $ spatial dimensions. Their
treatment has been done asymptotically in \cite{gsv}. We have
three general asymptotic behaviours:
\begin{itemize}
\item (i) {\bf unstable}  for $ R \to \infty $ in inflationary universes
with
$ \rho_u = u \, R^{2-D} ,\;   p_u =  -\rho_u/(D-1) < 0 $
\item (ii) {\bf Dual} to unstable  for $ R \to 0 $
with  $ \rho_d = d \, R^{-D},\; p_d = \rho_d/(D-1) > 0  $ .
\item (iii) {\bf Stable} for $ R \to \infty $,
with $\rho_s = s \, R^{1-D} , \;  p_s = 0 $ .
\end{itemize}

We find perfect fluid relations with the factors
\begin{equation}
\gamma_u = {{D-2}\over {D-1}} \quad , \quad \gamma_d =
  {D\over {D-1}} \quad , \quad\gamma_s = 1~.
\label{glud}
\end{equation}
This reproduces  the two dimensional and three dimensional results
for $ D = 2 $ and $ D = 3 $, respectively. The stable regime is absent
for $ D = 2 $ due to the lack of string transverse modes there.

The dual strings behave as {\it radiation} (massless particles) and the
stable strings are similar to {\it cold matter}. The unstable strings
correspond to the critical case of the so called {\it coasting universe}.
That is, classical strings provide a {\it concrete}
realization of such cosmological models.

Strings continuously evolve from one type of behaviour to another, as
is explicitly shown by our solutions \cite{prd,dls}. For
intermediate values of $ R $, the string equation of state is clearly
more complicated. We propose a formula of the type:
\begin{equation}
\rho = \left( u_R \; R + {{d} \over R} + s \right) {1 \over
{R^{D-1}}} ~~~,~~~
p  = {1 \over {D-1}} \left( {d \over R} -  u_R \; R\right) {1 \over
{R^{D-1}}} \label{rope}
\end{equation}
where
\begin{eqnarray}
\lim_{R\to\infty} u_R = \cases{ 0 \quad & {\rm FRW } \cr
  u_{\infty} \neq 0 & {\rm Inflationary } \cr}
\end{eqnarray}
is qualitatively
correct for all $ R $ and becomes exact for $ R \to 0 $ and $ R \to
\infty $. The parameter $u_R$ varies smoothly with $R$ and tends to
the constant $ u_{\infty}$ for $ R \to \infty$.

We stress here that we obtained the string equation of
state from the exact string evolution in cosmological spacetimes.

Inserting the equation of state (\ref{rope}) in the Einstein-Friedmann
equations of general relativity, we obtain a self-consistent solution
for $ R $ as a monotonically increasing function of the cosmic time $
T $
\begin{equation}
T = \sqrt{{(D-1)(D-2)}\over 2}~
\int_0^R dR \; {{R^{D/2-1}} ~\over {\sqrt{ u_R \; R^2  + d + s\; R}}}
\label{inte}
\end{equation}
where we set $R(0) = 0$.
This string dominated universe starts at $
T = 0 $ with a radiation dominated regime $
R(T) \buildrel{T \to 0}\over \simeq C_D \, (T)^{2 \over D}$,
then the universe expands for large $ T $ as
$ R(T) \buildrel{T \to \infty}\over \simeq C'_D ~
 (T)^{2 \over {D-1}}$, as  (cold) matter dominated universes.
For example, at $ D = 4 ,\;  R $ grows  as  $ \, R \sim   (T)^{2 \over 3}$.

It must be noticed that an universe dominated by unstable strings
($u_R$) would yield $ R(T) \buildrel{T \to \infty}\over \simeq C'_D ~
 (T)^{2 \over {D-2}}$, which is faster than (cold) matter
dominated universes. However, this is not a self-consistent solution
of the Einstein-Friedmann equations plus the string equations of
motion, as shown in \cite{cos}.

The unstable string solutions are called in this way since their
energy and invariant length grow as $R$ for large $R$. However, it
must be clear that as {\it classical} string solutions they {\bf never
decay}.

Our self-consistent solution $R(T)$ yields the realistic matter
behaviour $R(T)  \buildrel{T \to \infty}\over \sim (T)^{2 \over
{D-1}}$ .

The {\it stable} strings (which
behave as cold matter) are those dominating for $R \to \infty$. The
`dual' strings  give
$R(T) \buildrel{T \to 0}\over \simeq C_D \, (T)^{2 \over D}$,
 the radiation type behaviour.
For intermediate $R$, the three types of string behaviours (unstable,
dual and stable) are present. Their cosmological implications as well
as those associated with string decay deserve investigation. For a
thermodynamical gas of strings the temperature $T$ as a function of
$R$, scales as $1/R$ for small $R$ (the usual radiation behaviour).

For the sake of completeness we analyze the effective string
equations in \cite{cos}. These equations have been extensively
treated in the literature  and they are not our central aim.

It must be noticed that there is no satisfactory
derivation of inflation in the context of the effective string equations.
 This does not mean that string
theory is not compatible with inflation, but that the effective string
action approach {\it is not enough} to describe inflation. The
effective string equations are a low energy field theory approximation
to string theory containing only the {\it massless} string modes.
The vacuum energy scales to start inflation are typically of the order
of the Planck mass where the effective string action approximation
breaks down. One must also consider the {\it massive} string modes (which
are absent from  the effective string action) in order to properly get
the cosmological condensate yielding inflation.
De Sitter inflation does not emerge as a solution of the
the effective string equations.

\section{Conclusions}

Strings in curved spacetimes show a rich variety of new behaviours
unknown in flat spacetimes. The most spectacular effect is clearly given by
the unstable strings with size and  energy tending to infinity.
This phenomenon appears both for singular plane wave spacetimes and for
non-singular  de Sitter spacetimes and for all other inflationary
universes. We think that it is a generic feature for strings in strong
gravitational fields.

Comparison of  the energy-momentum behaviours in inflationary
universes and singular plane-waves show interesting differences.
We find for the fastest growing energy-momentum components in singular
plane-waves ($\b \geq 2$) \cite{ondpl}:
\begin{eqnarray}
T^{VV}(\s,\tau) \buildrel{\tau \to 0}\over= C_V \, [\xi(\tau)]^2
\int_0^{2\pi} [B^X(\s)]^2 \to \infty \cr
T^{VX}(\s,\tau) \buildrel{\tau \to 0}\over= C_X \, \xi(\tau)
\int_0^{2\pi} B^X(\s)  \to \infty
\end{eqnarray}
where
$$
\xi(\tau) \equiv |\tau|^{-\b/4}\; \exp[K|\tau|^{1-\b/2}] \to \infty
$$
and $ C_V $ and $ C_X $ are constants.

A typical unstable string behaviour on an inflationary spacetime with
scale factor $R(T) = a \; T^{{k \o {k+2}}} \; (k < 0)$ is as follows
\cite{eri94}.
\begin{eqnarray}
E  \buildrel{\tau \to 0}\over= {C \o {\a'}} \; \tau^{k/2} = {R \o
{\a'}}\to +\infty \cr
P \buildrel{\tau \to 0}\over=  -E/2 \to -\infty \cr
S  \buildrel{\tau \to 0}\over=\tau^{k/2}\to +\infty
\end{eqnarray}
where $C$ is a constant and we considered a ring solution for simplicity.

The main difference is that in singular plane wave spacetimes only the
null energy (conjugated to the null variable $V$) diverges. In the
inflationary case, both the energy ($T^0_0$) and the pressure
($-T^i_i$, not summed) diverge and they blow up at the same rate.
Obviously the kind of  spacetime singularity is very different.


\bigskip



\begin{thebibliography}{11}
\bibitem{eri92} H.J. de Vega and N. S\'{a}nchez in ``String Quantum
 Gravity and the Physics at the
              Planck Scale'', Proceedings of the Erice Workshop held in June
              1992. Edited by N. S\'{a}nchez, World Scientific, 1993.
              Pages 73-185, and references given therein.

\bibitem{eri94} H.J. de Vega and N. S\'{a}nchez in ``Current Topics
in Astrofundamental Physics'', Proceedings of the Chalonge School,
Erice  4-16 September 1994,
Proceedings edited by N. S\'{a}nchez and A. Zichichi.

\bibitem{dvs87} H. J. de Vega and N. S\'anchez, Phys. Lett. {\bf B 197}, 320
(1987).

\bibitem{negro} H. J. de Vega and N. S\'anchez,
	Nucl. Phys. {\bf B 309}, 552 and 577 (1988).

\bibitem{cos} H. J. de Vega and N. S\'anchez,
 Phys. Rev. {\bf D50}, 7202 (1994).


\bibitem{ondpl} H. J. de Vega and N. S\'anchez, Phys. Rev. {\bf D 45} ,
2783 (1992).

\bibitem{ondpl2} H. J. de Vega, M. Ram\'on Medrano and N. S\'anchez,

Class. and Quantum Grav. {\bf 10}, 2007 (1993).

\bibitem{camas} H. J. de Vega and N. S\'anchez,
Nucl. Phys. {\bf B 317},  706 and 731 (1989).

H. J. de Vega and N. S\'anchez, Phys. Rev. Lett. (C),  {\bf 65}, 1517 (1990).

\bibitem{horo} G. Horowitz and A.R. Steif, Phys. Rev. Lett. {\bf 64},
260 (1990)

and  Phys. Rev. {\bf D 42} , 1950 (1990).


\bibitem{guven} R. G\"uven, Phys. Lett. {\bf B 191}, 275 (1987).

 D. Amati and K. Klim\^cik,  Phys. Lett. {\bf B 210} , 92 (1988).

\bibitem{trans} H. J. de Vega, M. Ram\'on Medrano and N. S\'anchez,

Nucl. Phys. {\bf B 351}, 277 (1991),
 Nucl. Phys. {\bf B 374}, 425 (1992)

and  Phys. Lett. {\bf B 285}, 206 (1992).

\bibitem{prd} H. J. de Vega and N. S\'anchez,
Phys. Rev. {\bf D47}, 3394 (1993).

\bibitem{dms} H. J. de Vega, A. V. Mikhailov and N. S\'{a}nchez,

Teor. Mat. Fiz. {\bf 94} (1993) 232.

\bibitem{cdms} F. Combes, H. J. de Vega, A. V. Mikhailov and
N. S\'{a}nchez,

Phys. Rev. {\bf D50}, 2754 (1994).

\bibitem{dls} H. J. de Vega, A. L. Larsen and N. S\'anchez,
Nucl. Phys. {\bf B 427}, 643 (1994).

\bibitem{igor} I. Krichever, Funct. Anal. and Appl. {\bf 28}, 21 (1994),

[Funkts. Anal. Prilozhen.  {\bf 28}, 26 (1994)].

\bibitem{din} H. J. de Vega and I. L. Egusquiza,
Phys. Rev. {\bf D 49}, 763 (1994).

\bibitem{ads}  A. L. Larsen and N. S\'anchez,
Phys. Rev. {\bf D 50}, 7493 (1994).


\bibitem{sv} N. S\'anchez and G. Veneziano,
Nucl. Phys. {\bf B 333}, 253 (1990).

\bibitem{gsv} M. Gasperini, N. S\'anchez and G. Veneziano,

Int. J. Mod. Phys. {\bf A 6},  3853 (1991) and
Nucl. Phys. {\bf B364}, 365 (1991).

\bibitem{ijm} H. J. de Vega and N. S\'anchez,
Int. J. Mod. Phys. {\bf A 7}, 3043 (1992).

\end{thebibliography}
\end{document}